\documentclass[aps,prl,twocolumn,showpacs,superscriptaddress]{revtex4-1}  

\usepackage{graphicx}  
\usepackage[space]{grffile}
\usepackage{dcolumn}   
\usepackage{bm}        
\usepackage{amssymb}   
\usepackage{braket}
\usepackage{gensymb}
\usepackage{amssymb,amsmath,amsbsy}
\usepackage[utf8]{inputenc}
\usepackage{hyperref}
\usepackage[ampersand]{easylist}
\usepackage{xcolor}

\graphicspath{{figures/}}

\newcommand{\sample}{Er:Y$_2$SiO$_5$}

\begin{document}

\title{Optically unstable phase from ion-ion interactions in an erbium doped crystal}

\author{ Yu-Hui Chen}  \email{These authors contributed equally to this work. } 
\affiliation{School of Physics, Beijing Institute of Technology, 5 South Zhongguancun
Street,  Haidian District, Beijing 100081, China}
\author{ Sebastian P. Horvath} \email{These authors contributed equally to this work. }
\affiliation{Department of Physics, Lund University, P.O. Box 118, SE-22100 Lund, Sweden.}
\author{Jevon J. Longdell}
\affiliation{The Dodd-Walls Centre for Photonic and Quantum Technologies, Department of Physics, University of Otago, 730 Cumberland Street, Dunedin, New Zealand.}
\author{Xiangdong Zhang}
\affiliation{School of Physics, Beijing Institute of Technology, 5 South Zhongguancun
Street,  Haidian District, Beijing 100081, China}

\date{\today}

\begin{abstract}
Optical many-body systems naturally possess strong light-matter interactions and are thus of central importance for photonic applications. However, these applications are so far limited within the regime of intrinsic dynamically-stable phases, and the possibility of  unstable phases remains unidentified. Here we experimentally revealed a new dynamical phase of intrinsic optical instability by using a continuous-wave laser to drive an erbium doped crystal. The transmission through the sample became unstable for intense laser inputs, and transient net gain was observed if the light passed the sample twice. The phase transition, between states in and out of a dynamical equilibrium, was induced by the dipole-dipole interactions between nearby erbium ions.
\end{abstract}


\maketitle
 {Strong light-matter interaction is essential for photonic technologies. Due to collective many-body enhancement, atomic ensembles naturally possess strong light-matter interaction, and a recent surge of new photonic applications are based on the emergence of new optical phases \cite{Sun2017, Flick2018, Gorshkov2011}. For example, the chaotic behaviour of a laser is a resource for reservoir computing instead of a nuisance \cite{Kanno2012, Appeltant2011}, and novel coherent light sources can be achieved in a phase of polaritonic Bose-Einstein condensation \cite{Kasprzak2006}. These systems are genuinely nonequilibrium many-body systems, and due to atom-atom interactions rich phases arise even without external feedback. Mainly because of the complexity induced by optical driving, the interplay of atoms, and different kinds of losses,  clarifying new intrinsic optical phases is still challenging. So far all explored intrinsic phases of optical quantum systems are limited in the regime of dynamically stable phases, and the possibility of dynamical instabilities, which is the other half of the phase diagram, still lacks experimental evidence \cite{Vlasov2013}.}

{The phases of an atomic system driven by an optical field have been extensively studied for decades. Optical bistability in a passive system was first demonstrated by Gibbs using an optical cavity to provided feedback \cite{Gibbs1987}. It was soon realized that optical many-body systems also provided feedback and gave rise to mirrorless bistability or intrinsic optical bistability \cite{Carmichael1977, Bowden1979}. The first observation of this were demonstrated using semiconductors \cite{Zental1977, Bohnert1983, Hajo1983}. Here, feedback originates from an increase of the optical absorption as the excitation is increased \cite{Miller1984}. Another kind of intrinsic optical bistability originates from atomic many-body interactions, that is, the state of an atom affects the local electric field or magnetic field of neighboring atoms and determines whether the nearby atoms are resonant with an applied optical field \cite{Bowden1979, Diehl2010, Lee2011, Karabanov2017}. Intrinsic bistability arising from atom-atom interactions has been observed in rare-earth doped crystals and Rydberg ensembles  \cite{ Hehlen1994, Carr2013, Bloch2008,  Goldner2004}. However, the two possible outputs of bistability are still dynamically-stable states. Although it is well accepted that nonlinear quantum master equations allow for an unstable response, whether it really exists remains unclear to date \cite{Diehl2010, Vlasov2013}. A recent theory, which concludes that dynamical instabilities are absent for nondegenerate two-level homogeneous systems \cite{Vlasov2013}, further indicates that a very specific set of experimental conditions are required in order to observe this type of instability. The lack of experimental observation of intrinsic instabilities hinders applications of optical many-body systems in the regime of, for example, random number generation, transmission security, and chaotic optical sensing \cite{Sciamanna2015}.  }

Here we experimentally demonstrated a new optical phase of intrinsic instabilities using an erbium doped yttrium orthosilicate crystal (\sample). In our experiment we observed a dynamically unstable response when measuring  the light transmission of a strong continuous-wave laser input. Furthermore, if the optical path-length was increased by reflecting the light such that it passed the sample twice, then transient net gain was observed.

\begin{figure}
  \includegraphics[width=\columnwidth]{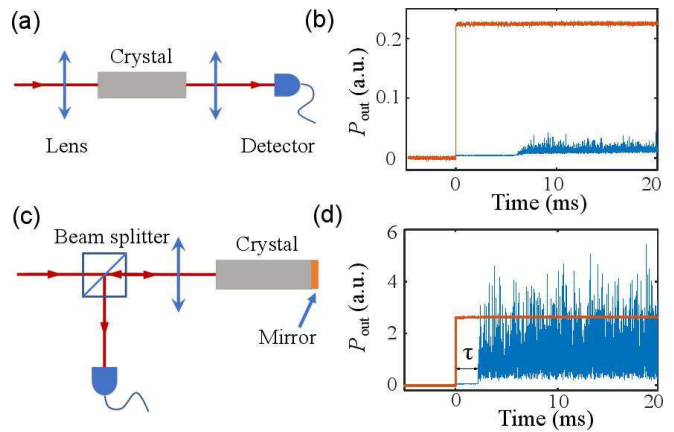}
  \caption{\label{fig.expsetup}
  Experimental setup and dynamical instabilities. (a) Schematic of the direct transmission measurement. A laser is incident from the left side and hits a detector on the right. (b) Measured transmitted signal corresponding to setup (a). The incident light is switched on at $t = 0$\,ms.  \emph{Orange:} transmission curve when the laser frequency is far from the erbium absorption, $f_l=16.53$\,GHz, and $P_{\text{in}}=27.9$\,mW; \emph{Blue:} transmission curve when the laser frequency is on resonance $f_l=-0.34$\,GHz, and $P_{\text{in}}=26.8$\,mW. (c) Schematic of the double-pass measurement. A beam splitter is used to guide the reflected light to our detector. (d) Measured reflected signal corresponding to setup (c). \emph{Orange:} data for off-resonance $f_l = 72.85$\,GHz and $P_{\text{in}} = 14.6$\,mW; \emph{Blue:} data for on-resonance $f_l = 0.41$\,GHz and $P_{\text{in}} = 14.6$\,mW.  Characteristic parameter $\tau$ is defined as the delay between the moment when the laser is switched on and the moment when instabilities arise.}
\end{figure}

The measurements were performed on two distinct configurations, as shown in Fig. \ref{fig.expsetup}, which consisted of a direct transmission and a double-pass measurement.  Two cylindrical \sample \ samples (from Scientific Materials), which had the same diameter of 5\,mm and length of 12\,mm, were placed in a homemade cryostat and cooled to 4.2\,K. The samples had 1000\,ppm of yttrium ions replaced by erbium ions. For the direct transmission measurement, neither surface of the sample was coated. Two lenses with a focal length of 300\, mm were placed in a confocal configuration outside the cryostat, as shown in Fig. \ref{fig.expsetup}(a). For the double-pass measurements, a single lens with a focal length of 250\,mm was used to guide light to the sample, as shown in Fig. \ref{fig.expsetup}(c). One end of the sample was prepared with a convex shape of $r=$100.305\,mm curvature and was coated to achieve a reflectivity of 98.8\%, and the other end was flat and anti-reflection coated. In both setups,  the radius of the focal spot was approximately 100 $\mu$m; therefore a $\mathcal{O}$(1\,mW) laser input  corresponded to a $\mathcal{O}$(0.1\,MHz) Rabi frequency. The outgoing beams were well collimated (with a radius of about 1.5\,mm ), and a lens with a focal length of 50\,mm was used to collect light onto a photo detector with a bandwidth of  $300$\,MHz. The laser used in our experiment was a commercial laser (Adjustik Koheras, NTK) with a linewidth of less than 1\, kHz.

The studied resonance frequency of \sample \ was at 1536.4\,nm (195117.17\,GHz), corresponding to spectroscopic site 1 (following the literature convention\cite{Maksimov1970}). For notational convenience, all quoted laser frequencies $f_l$ are relative frequencies, that is,  the absolute frequency minus 195117.17\,GHz. For the direct transmission measurements, if the laser frequency $f_l$ was off resonance from the erbium absorption, the detected signal when the laser was switched on was, as expected, a step function ($t = 0$\,ms, orange line in Fig. \ref{fig.expsetup}(b)). However, as $f_l$ was tuned on resonance, a threshold power could be reached beyond which the detected signal became dynamically unstable, which is shown by the blue line in Fig. \ref{fig.expsetup}(b). 
The observed instability was considerably more obvious if the light was reflected back and passed the sample twice, as is shown in Fig. \ref{fig.expsetup}(c) and (d).
The measured reflection signal (blue line in Fig. \ref{fig.expsetup}(d)) became unstable after a delay $\tau = 2.2$\,ms and occasionally jumped above the off-resonant reflection (orange line). This means that for some specific instances in time the output power of the unstable state was stronger than its input (detailed in Section V.A, Supplementary Information(SI) \cite{SI2021}).

\begin{figure}[!]
  \includegraphics[width=\columnwidth]{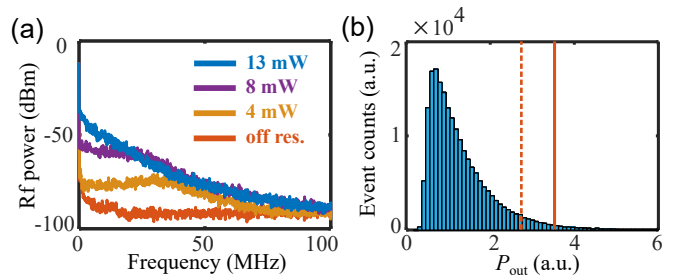}
  \caption{\label{fig.stat}
  Data analysis of the double-pass experiment. (a) Spectral analysis of signals from the detector. \emph{Orange:}, off resonance, $f_{l}=22.93$\,GHz, $P_{\text{in}} =12.9$\,mW. On resonance spectra of $f_{l}=0.45$\,GHz were measured for different laser inputs as noted.  (b) Stochastic intensity distribution for $f_{l}=0.45$\,GHz and $P_{\text{in}} =12.9$\,mW. The sampling interval is 250\,ns, and the total sampling time is 200\,ms. The $x$ axis is the voltage values from our detector, and the $y$ axis is the event counts. The orange dashed line is the estimated input for the blue distribution.  The orange solid line shows the distribution of the output of an off-resonance laser $f_{l}=21.03$\,GHz and $P_{\text{in}} =12.9$\,mW. }
\end{figure}

Figure \ref{fig.stat}(a) shows the power-spectral density of the unstable signal, demonstrating significant frequency components of bandwidth $\mathcal{O}$(10\,MHz).  For the double-pass experiments,  when $f_l$ was far from the erbium absorption, there was no instabilities and the corresponding spectrum featured a sharp peak near zero-frequency. When $f_l$ was on resonance and the input power was increased, the detected signal became unstable and the spectra were broadened with cut-off response frequencies of $\sim$50\,MHz. In contrast, the noise of the laser has a bandwidth less than $1$\,kHz. The low frequency response in the range 0-20\,kHz can be found in Section IV, SI \cite{SI2021}.  Shown in Fig. \ref{fig.stat}(b) is a stochastic distribution of the unstable output $P_{\text{out}}$. The orange dashed line shows the estimated input power. Instead of being a $\delta$ function corresponding to a constant output, the unstable data shows a broad distribution with a maximum likelihood located at $P_{\text{out}} = 0.7$. From the tail of the distribution, one can tell that there was a chance that the detected signal surpassed the input, which agrees with Fig. \ref{fig.expsetup}(d).

\begin{figure}
  \includegraphics[width=\columnwidth]{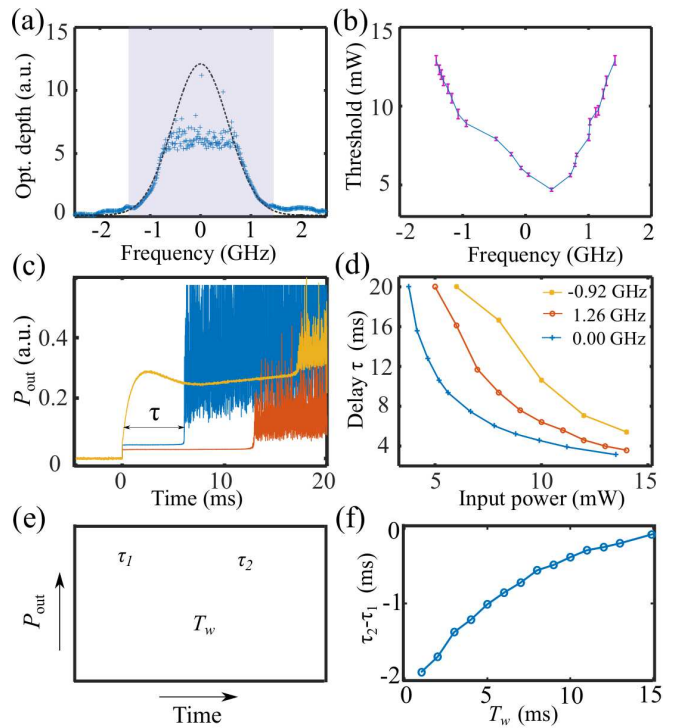}
  \caption{\label{fig.depend}
  Dependencies on input power and frequency, for the double-pass experiment. (a) Optical absorption of the sample. Cross points: measured data using low- power input. The signal near the absorption center became less measurable due to the large optical depth and the low input power (high input power would generate instabilities); the derived optical depth is therefore noisy. Dashed line: a Gaussian fit. The shaded area marks the frequency range where we could observe instabilities. 
(b) Power Threshold as a function of laser frequency. (c) Typical measured data of different laser powers and frequencies. \emph{Blue:} $f_l = 0.00$\,GHz, $P_{\text{in}}=7.8$\,mW (note: here the detector was not saturated, but the y-axis was clipped for clarity of the remaining data); \emph{Orange:} $f_l = 0.00$\,GHz, $P_{\text{in}}=4.7$\,mW; \emph{Yellow:} $f_l = -1.39$\,GHz, $P_{\text{in}}=7.8$\,mW. (d) Delay $\tau$ as a function of $P_{\text{in}}$ for different $f_l$ as noted. (e) Schematic of input laser sequence. The duration of the two driving pulses is 20\,ms. The $\tau$ of the two pulses is written as $\tau_1$ and $\tau_2$, and the waiting time between the pulses is $T_{w}$. (f) Measured $\tau_2-\tau_1$ as a function of $T_w$. Fitting of the data shows a time constant of 13.8\,ms. $f_{l} = 7.64$\,GHz and $ P_{\text{in}} = 13.3$\,mW.
  }
\end{figure}

Figure \ref{fig.depend} shows the dependence of the instabilities on the laser frequency $f_l$ and the input power $P_{\text{in}}$. Figure \ref{fig.depend}(a) are the measured and fit inhomogeneous absorption lines of our sample. Over the range of the inhomogeneous line, the threshold of the instabilities were measured, that is, the minima of $P_{\text{in}}$ for the instabilities to occur at a given $f_l$, as shown in Fig. \ref{fig.depend}(b). Within the power limit of our laser, the range within which instabilities could be observed is $-1.57 \le f_l \le 1.45$\,GHz, and matches the inhomogeneous linewidth of the erbium ions, as shown in Fig. \ref{fig.depend}(a) and (b). Due to the different spatial density of the erbium ions across the inhomogeneous line, scanning the laser frequency across the inhomogeneous line results in different excitation densities. This emphasizes the role of excitation-induced ion-ion interactions. Measurements on our 50\,ppm and 10\,ppm samples manifested no instabilities, also indicating that weak ion-ion interactions can not generate the instabilities (excited ion density is not directly proportional to optical density, see Section V.D, SI \cite{SI2021}). By applying a magnetic field to our sample, the absorption peak in Fig. \ref{fig.expsetup}(a) can be Zeeman split into four peaks, two of which had an optical depth $>5$. However, no instabilities were recorded in any of these peaks (Section V.B, SI \cite{SI2021}). In contrast, in Fig. 3(a) and (b) instabilities arose even for low optical depth $\sim$ 1. This is expected. Once the energy levels are split, the erbium ions become a non-degenerate two-level system, in which, according to the Bloch equations, dynamical instabilities are not expected regardless of the driving field and losses \citep{Vlasov2013}.  

In addition, the dependencies on $f_l$ and $P_{\text{in}}$ indicates an accumulating effect of the excited ions. As shown in Fig. \ref{fig.depend}(c) and (d), depending on $f_l$ and $P_{\text{in}}$, it took a varying amount of time ($\tau$) to establish an instability. For a strong $P_{\text{in}}$  less time was needed. More evidence is shown in Fig. \ref{fig.depend}(e) and (f). In the experiment, two subsequent laser pulses were input to the sample; between the two pulses there was a waiting time $T_w$ to allow the excited ions to repopulate. If $T_w$ was short and the ions excited by the first pulse remained unchanged, the second pulse took  less time to establish an instability. $\tau_2$ and $\tau_1$ of the two pulses thus showed a difference that depended on $T_w$. It took about $\sim$10\,ms to eliminate the effect of the first pulse, which is consistent with the 11\,ms lifetime of the excited state.

\begin{figure}
  \includegraphics[width=\columnwidth]{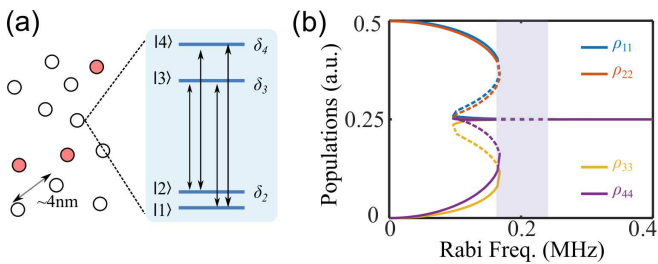}
  \caption{\label{fig.model}
  A model for the instabilities. (a) Microscopic description and energy level configuration. Ions that sit in their ground (excited) states are symbolized by open (filled) circles. The average distance of randomly distributed site-1 erbium atoms is approximately 4\,nm.  Without applied magnetic field, the degeneracies of both ground states and excited states are assumed to be split by the erbium-erbium interactions. (b) Calculated populations of the four levels for increasing driving Rabi frequency. $\rho_{11}$, $\rho_{22}$, $\rho_{33}$, and $\rho_{44}$ are indicated by different colours as noted. The data with dashed-lines are associated with positive eigenvalues and are thus unstable. The intrinsic-instability regime is indicated by the shaded area. The Rabi frequency used in calculations agrees with our experiment conditions.}
\end{figure}

These experimental results provide some insights into the physics behind the observed instabilities. Given the frequency stability of the laser, as well as the thermomechanical stability of the apparatus, the response bandwidth spanning several 10s of MHz indicates that the instabilities originate from ion-ion interactions. Other alternative explainations that were ruled out are enumerated and discussesed in Section III, SI \cite{SI2021}. 
 
Considering in closer detail the ion-ion interactions, rare-earth ions normally possess magnetic and/or electric dipole moments (Section I, SI \cite{SI2021}). Because these solid-state materials feature extremely long optical and spin coherence times \cite{Sun2002, Zhong2015}, the dipole-dipole interactions between nearby ions \cite{BIRGENEAU1969} is experimentally distinguishable from various decoherence processes. Optical excitation of an erbium ion can instantaneously change the local electric or magnetic field of neighboring ions, and shift their optical-frequencies as a result of the Stark effect or the Zeeman effect \cite{Huang1989, Liu1990}. This excitation induced frequency shift agrees well with the experimental results of concentration-dependent \cite{Liao1973, Taylor1974} and intensity-dependent \cite{Huang1989} photon-echo relaxation rates, and is also referred to as instantaneous spectral diffusion \cite{Huang1989, Liu1990}. In addition, optically-pumped spins can gradually polarize neighbouring spins through ground-state flip-flop interactions, and thus change the local magnetic field. This can, similarly to the above discussed excitation dependent effect, shift the frequency of ions, but would only require the pumping of a small fraction of spins \cite{Welinski2020}.  For many applications, the excitation-induced effect is considered to affect the decoherence of a system and is simplified as an additional dephasing term. Further investigations have shown that it can also introduce feedback to the system and give rise to an intrinsic-optical-bistability phase \cite{Bowden1993, Guillot-Noel2002, Kuznetsov2011}.

Here we theoretically show that such a frequency shift can also result in a dynamically unstable phase. To obtain a basic understanding we applied the mean-field approximation (Section I, SI \cite{SI2021}) and disregarded the inhomogeneities (Section V.C, SI \cite{SI2021}). The effects of ion-ion interactions are : (1) the spin states are slightly split by an average magnetic field produced by all nearby ions, that is, we assume the frequency differences between the two ground state and between the two excited states are $\mathcal{O}$(0.1\,MHz), which makes erbium ions a nearly-degenerate system with four levels, as illustrated in Fig. \ref{fig.model}(a); (2) the resonant frequency of the $i$th ion will be changed if its nearby ions are optically excited. Thus the Hamiltonian of an erbium ion is written as
\begin{equation} \label{eq.ham}
\begin{split}
& H = H_0 + H_f + H_d,\\
& H_0 = \delta_{2} \sigma_{22}+\delta_{3} \sigma_{33}+\delta_{4} \sigma_{44}, \\
& H_f = \Omega_a (\sigma_{31} +\sigma_{42}) + \Omega_b (\sigma_{41} + \sigma_{32}),\\
& H_d = \Delta_s (\rho_{44}+\rho_{33}-\rho_{22}-\rho_{11})(\sigma_{33} + \sigma_{44}),
\end{split}
\end{equation}
where $H_0$ is the Hamiltonian of a free ion, $H_f$ is the interaction with a laser field, $H_d$ stands for the dipole-dipole interactions \cite{Graf1998},  $\delta_i$ is the detuning of the $i$th level, $\sigma_{ij} \equiv |i \rangle \langle j |$, $\Omega_a$ and $\Omega_b$ are the driving Rabi frequencies due to different oscillator strengths, $\rho_{ii}$ are the diagonal elements of the mean-field density matrix of a single ion, and $\Delta_s$ is the excitation induced frequency shift.

Using the Hamiltonian Eq.(\ref{eq.ham}) and introducing decay and dephasing terms into the Lindblad master equation enables us to calculate the response of our system. Shown in Fig. \ref{fig.model}(b) is a calculation of the populations of the four-level system using typical parameters of erbium atoms at a temperature of 4\,K. An intrinsic-optical-bistability regime exists where there are multiple population solutions for one laser input. What is more, there is a regime where the steady-state solution corresponding to a highly-populated excited state becomes unstable, indicated by the shaded area in Fig. \ref{fig.model}(b). Because an absorption measurement is directly proportional to the population difference between the ground  and the excited states, an unstable $\rho_{ii}$ manifests itself as an unstable absorption and the measured transmission or reflection thus becomes unpredictable as was observed in our experiment.

We note that it is $H_d$ that introduces nonlinearity to the many-body system and gives rise to the unstable phase. A non-negligible $H_d$ requires a significant population in the excited states and a large $\Delta_s$. Erbium ions in this host are easy to saturate because of their long (11\,ms) excited state lifetime. This means even a modest laser input of,  13\,mW, can saturate over 100 MHz of the inhomogeneous line (see Section V.D, SI \cite{SI2021}).

$\Delta_s$ stems from the local-field variance due to the change of optical or spin states of nearby ions. This ion-ion interaction can only be seen if ions are spatially close enough, otherwise the frequency shifts would be buried in the background consisting of various decoherence processes.  For our 1000 ppm \sample \ crystal, the distance between two nearby erbium ions is about 4\,nm. At this distance, erbium ions with electron spins cause a $\mathcal{O}$(10\,MHz) magnetic dipole-dipole interaction (Section I.A, SI \cite{SI2021}). In addition, erbium ions also possess electric dipole moments, but to the authors' knowledge the full anisotropic Stark-shift of  \sample \ at 1.5\,$\mu$m is still to be measured. Noting that Er$^{3+}$ ions in different hosts have Stark coefficients typically between 10 and 100 $\text{kHz} \cdot \text{V}^{-1} \text{cm}$ \cite{Hastings-Simon2006,Craiciu2020}, we estimated the dipole induced Stark-shift of our system is  $\mathcal{O}$(10\,MHz) (Section I.A, SI \cite{SI2021}). Thus contributions of the magnetic and the electric interactions to $\Delta_s$ are likely of a comparable magnitude.

The effect of $H_d$ on the macroscopic polarization that is initially produced by light can be understood in the same way as phase modulation. Such a modulation arises from the random distribution of optical excitation and spontaneous emission, and thus makes the system unpredictable. A large modulation depth $\Delta_s$ $\mathcal{O}$(10\,MHz) generates a frequency response of similar order, which agrees with Fig. \ref{fig.stat}(a) (see Section IV, SI \cite{SI2021}). The mediated polarization remains coherent with the driving laser until the ions decohere, therefore the instability can last for several $\mu$s even after the input light is switched off (Section IV, SI \cite{SI2021}). If the transmitted light, which is already unstable, is reflected back to the sample within the excited-state coherence time, the ensemble behaves as a loss or gain medium depending on its history and the light gets absorbed more or amplified intensely. 


In summary, we have demonstrated a dynamically unstable phase by using a continuous-wave laser to drive a dense erbium-doped crystal. Due to the ion-ion interactions between nearby erbium ions, optically exciting an erbium ion can bring the nearby ions into resonance or out of resonance with the laser. Such an effect is usually considered as an extra source of dephasing and is known as one origin of intrinsic optical bistabilities. Our experiment shows that, for a non-two-level system, it can also bring the system to a regime that is out of dynamical equilibrium. The underlying mechanism of the observed instabilities is closely related to the photon blockade  in Rydberg atoms \cite{Lukin2001} and such a phase is of particular relevance for quantum gate-operation schemes based on the ion-ion interactions in stoichiometric rare-earth ion crystals \cite{Ahlefeldt2013}. Beyond this, the developed model is sufficiently general, and thus consists of an interesting material to utilize the laser-driven many-body system for photonic applications. For example, combined with an optical cavity, the instabilities have potential applications in developing novel chaotic light sources and chaos-assisted on-chip frequency combs \cite{Chen2020}.


\begin{acknowledgments}

The authors would like to thank Howard Carmichael for helpful discussions, and Haitham El-Ella for helpful discussions and for sharing some preliminary modelling work. The authors wish to acknowledge financial support from the Marsden Fund of the Royal Society of New Zealand through Contract No. UOO1520,  the Start-up Fund of Beijing Institute of Technology, and the Science and Technology Innovation Project of Beijing Institute of Technology.

YHC and SPH contributed equally to this work.
\end{acknowledgments}

%


\end{document}